\documentclass[letterpaper, final]{aipproc}

\layoutstyle{8x11double}
\pdfpagewidth=\paperwidth
\pdfpageheight=\paperheight

\newcommand{\demographicstable}{
\begin{table}
\begin{tabular}{lcc} \hline
& \tablehead{1}{l}{b}{Compass~(\%)}
& \tablehead{1}{l}{b}{Department~(\%)} \\ \hline
Female & 47 & 16 \\
Chicano/Latino & 21 & 7 \\
Black & 5 & 0.5 \\
Native American & 3 & 0.5 \\
First-generation & 26 & N/A \\ \hline
\end{tabular}
\caption{Demographic data for the 2008-2011 Compass summer program students ($N=62$) versus UC Berkeley Physics Department averages.}
\label{tab:demographics}
\end{table}
}

\newcommand{\academicyeartable}{
\begin{table}
\begin{tabular}{lcc} \hline
& \tablehead{1}{l}{b}{Community~(\%)}
& \tablehead{1}{l}{b}{Organizers~(\%)} \\ \hline
Freshmen & 19 & 23 \\
Sophomores & 16 & 13 \\
Juniors & 13 & 11 \\
Seniors & 15 & 11 \\
Graduate Students & 27 & 37 \\
Alumni & 10 & 5 \\\hline
\end{tabular}
\caption{Composition of Compass community ($N=108$) and organizers ($N=38$) during 2011/12 academic year.}
\label{tab:academicstatus}
\end{table}
}

\newcommand{\quotetable}{
\begin{table}
\renewcommand{\arraystretch}{1.3}
\begin{tabular}{p{0.45\textwidth}} \hline
\tablehead{1}{p{0.45\textwidth}}{b}{After your first year as a Cal and Compass student, what (if anything) is important for you about Compass?}\\ \hline
"The skills you learn . . . Compass puts you in an environment where you really have to learn how to think with a science-y brain. Also the friends you make. Almost every single one of my friends at Cal I met through Compass."\\
"Compass has given me a community of fellow students and graduate students who I can go to whether I need help with my homework or just someone to talk to. Having Compass and its community as a first year student at a large university helped me find my way and transition to Cal." \\
"It's a great introduction to college and to physical sciences in college. It's good to get to know people before the first semester starts and settle down a little. It also helps to encourage students towards the physical sciences and towards research." \\
"I love the sense of community and belonging Compass offers. I always feel included in any activity and the people are very approachable. Compass has become a family for me and a support system I can count on for guidance and help." \\
"Helps me know that I am not the only person who finds difficulties with the physics and (more generally) undergraduate education at Berkeley. It gives me confidence that I can face the challenges in the physics education and keeps my eyes open about the fact that school is not everything in one's undergraduate education."\\
\hline
\end{tabular}
\caption{Typical responses to open-ended survey question.}
\label{tab:quotes}
\end{table}
}

\begin{document}

\title{Building Classroom and Organizational Structure Around Positive Cultural Values}


\author{Badr F. Albanna}{
  address={Department of Physics, University of California, Berkeley, California 94720}
}

\author{Joel C. Corbo}{
  address={Department of Physics, University of California, Berkeley, California 94720}
}

\author{Dimitri R. Dounas-Frazer}{
  address={Department of Physics, University of California, Berkeley, California 94720}
}

\author{Angela Little}{
  address={Graduate Group in Math and Science Education, University of California, Berkeley, California 94720}
}

\author{Anna M. Zaniewski}{
  address={Department of Physics, University of California, Berkeley, California 94720}
}

\begin{abstract}
The Compass Project is a self-formed group of graduate and undergraduate students in the physical sciences at UC Berkeley. Our goals are to improve undergraduate physics education, provide opportunities for professional development, and increase retention of students--especially those from populations typically underrepresented in the physical sciences. Compass fosters a diverse, collaborative student community by providing a wide range of services, including a summer program and fall/spring seminar courses. We describe Compass's cultural values, discuss how community members are introduced to and help shape those values, and demonstrate how a single set of values informs the structure of both our classroom and organization. We emphasize that all members of the Compass community participate in, and benefit from, our cultural values, regardless of status as student, teacher, or otherwise.
\end{abstract}

\maketitle


UC Berkeley is a top-tier research university that attracts a diverse student body, and its physics department is one of the largest in the country. Given its size, integrating with the community and developing a physics network can be a daunting task for incoming students, potentially affecting their grades~\cite{Gonyea2008}, mental health~\cite{Lustig2006}, and persistence in their field of study~\cite{Seymour1995}. Common student concerns that contribute to attrition in the sciences~\cite{Seymour1995} can be mitigated by student engagement in educationally purposeful activities~\cite{Gonyea2008} or taking classes that emphasize social aspects of learning like group work~\cite{Treisman1992}. In general, a positive culture and supportive community are important for making students feel like a part of a department~\cite{Duncombe2003}. We describe how the Compass Project~\cite{website}, a student-run program at UC Berkeley, creates such culture and community in and out of the classroom.

Founded in 2006, Compass is an organization whose goals are to improve undergraduate physics education, provide opportunities for professional development, and increase retention of students. Leadership and teaching roles in Compass are filled by graduate and undergraduate students--including the authors of this paper--and its services exist as optional supplements to the required curriculum. Women, minority, and first generation college students are especially encouraged to participate, so these students are more represented in Compass than in the physics department as a whole~(Table~\ref{tab:demographics}). Compass fosters a diverse, collaborative student community by providing a wide range of programs designed to support students holistically.  Our multifaceted approach simultaneously develops disciplinary, intellectual, and emotional skills for student success through a summer program, fall and spring courses, a mentoring program, and other forms of academic, social, and personal support.

\demographicstable

In this paper, we highlight four cultural values that are central to the Compass community: \emph{iterative problem-solving}, \emph{collaboration}, \emph{ownership}, and \emph{supporting the whole person}.  The Compass approach to problem solving is iterative:  we try a solution, observe the results, and use those observations to improve our solution.  In general, we develop and implement these solutions as a group while maintaining individual pride over our work, thus balancing the dual values of collaboration and personal ownership.  Lastly, collaborating on difficult projects forges strong friendships, ultimately creating an environment in which each person is supported by the community personally as well as academically, \emph{i.e.}, as a whole person.  We explore these four values in the dual contexts of the Compass classroom and organization, and present survey data that focuses on the value of supporting the whole person.


\section{The Compass Classroom}

We begin with a broad overview of the Compass classroom before giving more detailed examples of how our values inform our curricula and final projects. Compass offers a sequence of three courses: a summer program, a fall course, and a spring course. While the summer program has been offered since 2007, the fall and spring courses were added in 2009 and 2012, respectively.

Compass courses are co-taught by teams of teachers, typically pairing veteran teachers with new ones. All teachers but one have been graduate students; one teacher was a senior undergraduate. Curricula are structured around over-arching physics topics which students explore by engaging in student-driven small-group discussions and experiments.  Traditional lecturing is avoided; instead, teachers guide discussion by asking questions of students in the manner of Modeling Discourse Management~\cite{Desbien2002} and Think-Pair-Share~\cite{Lyman1981}. During discussions, teachers reinforce the appropriate use of conceptual and epistemological resources~\cite{Hammer2000} and encourage metacognition~\cite{Redish2003}. One particularly unique feature of our courses is that they serve dual purposes, namely, developing physics disciplinary skills and reinforcing behaviors that are beneficial to college success. Students and teachers are supported in achieving these goals by embedding the courses in a larger support network that includes mentoring, tutoring, undergraduate-oriented research colloquia, and opportunities for leadership.

The summer program is offered to 15-20 incoming freshmen at no cost to the student.  It is a two-week long residential program that incorporates both classroom and extracurricular activities.  Its  curriculum is designed and taught by a team of 6 graduate students, about half of whom are new to Compass.  The goals of the summer program are to establish strong friendships and group work skills among the undergraduates, to provide an opportunity for graduate students to develop pedagogical skills, and to fold new undergraduate and graduate students into the Compass community.

Experienced teachers introduce new ones to Compass's teaching pedagogy by collaborating on novel curricula for the summer program. Summer curricula are characterized by an overarching theme which changes every year. Past themes include earthquakes, special relativity, quantum mechanics, wind turbines, and non-Newtonian fluids; see Ref.~\cite{Roth2012} for an example of a typical curriculum.  The curriculum is divided into three classes: experimental, theoretical, and metacognitive, each of which is taught by a pair of teachers\footnote{In 2009 and 2011, the summer programs were only one week long, there were fewer teachers, and there was no metacognitive class.}.

Outside the classroom, students enjoy a full schedule of activities. Field trips, tours of research laboratories, and a scavenger hunt serve to build community among the incoming freshman and to familiarize them with the campus and department. In addition, students are paired with graduate student mentors with whom they will continue to meet regularly for the duration of their freshman year, and potentially thereafter.

After completing the summer program, participants are enrolled in Compass's fall and spring semester courses. Each course meets once per week for two hours and is taught by a pair of teachers. In the fall, students learn about physical models in the context of the ray model of light. This course culminates in an independent research project where students explore and develop a model to answer a physical question of their choice. In the spring, thermal expansion is the paradigm phenomenon for teaching and learning about scientific measurement. Helping students acclimate to college life and develop productive ways of interpreting grades are also major goals of Compass's semester courses.

Compass's values are explicitly built into the structure of our courses, as the following examples demonstrate.

In the fall course, students have \emph{ownership} over their final research project. This project allows them to explore a question of their choice that is personally interesting to them. To answer their questions, students develop models by talking with peers, consulting the literature, and conducting experiments. Graduate students are assigned to groups of students with similar topics, acting as advisors to help guide the research process. The last five weeks of class are spent supporting students' projects. At the end of the semester, students present their work to their classmates and the larger Compass community.

In the spring course, the students explicitly practice \emph{collaboration} and \emph{iteration} through the refinement of a thermal expansion experiment.  The apparatus for the experiment was designed by the teachers of the course, but the students use a consensus-based process to create the procedure that they will use to measure the expansion of a wire.  After the first round of measurement, students work together to improve the experiment by making changes to the apparatus, procedure, or data analysis methods. Students then make additional measurements, ultimately evaluating whether their changes are indeed improvements.

In both the fall and spring, students are \emph{supported as whole people} through the implementation of non-traditional homework assignments and class discussions. Typical homework assignments include attending office hours and meeting with the department's student support staff. In class and in online forums, students interpret grades as "academic measurements" and discuss assigned readings about situational and psychological effects that may affect their grades, \emph{e.g.}, stereotype threat~\cite{Mateer2008}, self-handicapping~\cite{Debus2003}, and motivation~\cite{Deci1984}.

Making sense of grades is especially important given that many of our students, used to getting good grades in high school,  lack productive strategies for dealing with grades that are lower than what they have come to expect.  As of spring 2012, students submit a weekly reflection in which they evaluate their behavior in one of their traditional math or science courses using a rubric that includes skills like persistence, intellectual courage, and mental resourcefulness. Self-evaluations help students track their growth using multi-dimensional, holistic measures that are useful complements to grades.


\section{The Compass Organization}

We continue with a description of Compass as an organization followed by specific examples connecting our values to our design and decision-making processes.

Graduate and undergraduate students are responsible for the design of Compass's programs and the work that makes those programs possible.  Besides a few paid teaching and evaluation positions, Compass is run on a volunteer basis with thousands of hours of work contributed each year.  The majority of organizers were involved with Compass before taking on a leadership role, either as students, teachers, or mentors. "Leadership roles" include coordinating the logistics of the summer program, evaluating Compass's effectiveness, planning social events, fund raising, and so on. Each year, Compass introduces about 30 new people into the community, some of whom inevitably take on leadership roles. In particular, undergraduates have taken on leadership positions to the point where the composition of the organization reflects the composition of the community served (Table \ref{tab:academicstatus}). This allows Compass to be a responsive, dynamic organization as new people propose new ideas for growth and improvement.

\academicyeartable

Compass's decision-making process is flat and consensus-based.  By "flat," we mean that any member of the community can attend any meeting and participate in whatever decisions are being made, regardless of academic year or length of time in Compass.  By "consensus-based," we mean that everyone present for a decision must consent to a proposed course of action for it to be accepted.  The process by which consensus is reached in the classroom is modeled after the process used by the organizers.

Compass's values play an important role in the organization, as we illustrate using the example of the evolution of the summer program.

Compass has hosted five \emph{iterations} of the summer programs to date. Each one is an improvement on the last, resulting in several positive changes to its schedule and "personality." In the first summer program, students spent upwards of eight hours per day in the classroom and had homework assignments late into the evening, leading to students feeling overworked. Contrast this to the current program, which has only five hours of class per day and allots time for community-building activities, such as a liquid nitrogen ice cream social, a dance social, and a scavenger hunt around campus. These changes occurred over the course of five years as Compass learned how to balance the summer program's dual goals of academic preparation and community building.

Graduate and undergraduate students \emph{collaborate} with one another and share \emph{ownership} over the summer program. The evolution of the program is facilitated by undergraduate leadership in Compass: students who went through the program as freshmen participate as coordinators and use their first-hand experience to improve it.   In particular, undergraduates interview teacher applicants, give input into the program's theme, recruit new Compass students, help with fundraising, and coordinate the logistics of the program. This direct involvement of undergraduates in the design process allows for more efficient and effective improvements than could be made by graduate students working alone.  Additionally, undergraduate involvement in leadership provides opportunities for students to develop important professional skills like grant writing.

Undergraduate leadership--and hence \emph{ownership} over Compass--extends beyond the summer program as well. In their capacity as leaders, undergraduates have played an important role in shaping all aspects of the organization, including Compass's myriad courses. For instance, the spring semester course was pioneered by a graduate and undergraduate student pair.  Combining the expertise and perspectives of graduate and undergraduate students in this way better enables Compass to serve its community because the people who are in tune with the needs of their immediate peers are empowered to create or modify services to address those needs. Shared leadership is therefore an important part of \emph{supporting the whole person}, a value that is clearly resonating with Compass students, as we show in the next section.


\section{Student Survey and Discussion}

In fall 2011, an online survey was administered to the students who participated in the 2010 summer program and the subsequent fall course. This group of students did not take a spring course because it was not offered that year. Of the 17 participants, 15 completed the survey.

The survey data suggest that these students felt \emph{supported as whole people} by the Compass community. Individual students' responses indicated that they found both academic and personal support critical to their freshman year.  In total, 14 students answered the survey question in Table~\ref{tab:quotes}; of those, 12 students spoke about community, friendship, or meeting people, and 6 students spoke about academic or scientific support. The table gives a sense of the range of student responses.

\quotetable

Multiple-choice questions from the survey provide evidence of strong, multifaceted relationships among Compass students. When asked about their interactions with other students from their cohort, 13 students said they saw each other at least once per week, 9 of whom saw each other daily. In addition, 8 studied together and 9 lived with, or planned to live with, another Compass student. Taken together, the frequency and nature of student interactions and their open-ended responses suggest positive, supportive  interpersonal relationships among the surveyed students.

Members of Compass forge strong bonds with each other by working together on challenging problems--whether scientific problems like modeling earthquakes, organizational ones like coordinating a summer program, or personal ones like making sense  of their grades. Building on the collective experience of our community leads to dynamic classroom and organizational structures that evolve in tandem with our understanding of our needs. By blending the roles of teacher, student, colleague, mentor, and friend, we are better able to help each other persevere and succeed at Berkeley. Most importantly, all members of our community participate in, and benefit from, our cultural values.

\begin{theacknowledgments}
We are grateful to B. Sadoulet; J. Bender; and G. Quan, N. Samuels, P. Merlo, and R. Nurmela for their advocacy, resource-sharing, and feedback, respectively. Since 2006, Compass has received support from private donations, NSF, and a number of sources at UC Berkeley. 
DRDF and AMZ were supported by NSF under grants PHY-1068875 and EEC-0832819, respectively.
\end{theacknowledgments}


\end{document}